\documentclass{iopjournal}
\usepackage{amsmath}
\usepackage{multirow}
\usepackage{amssymb}

\begin{document}

\articletype{Paper} 

\title{Template-Free Gravitational Wave Detection with CWT-LSTM Autoencoders: A Case Study of Run-Dependent Calibration Effects in LIGO Data}

\author{Jericho Cain$^{1,*}$\orcid{0000-0003-4731-9142}}

\affil{$^1$Physics Department, Portland Community College, Portland, OR, USA}

\affil{$^*$Author to whom any correspondence should be addressed.}

\email{jericho.cain@gmail.com}


\begin{abstract}
Gravitational wave detection requires sophisticated signal processing to identify weak astrophysical signals buried in instrumental noise. Traditional matched filtering approaches face computational challenges with diverse signal morphologies and non-stationary noise. This work presents an unsupervised deep learning methodology integrating Continuous Wavelet Transform (CWT) preprocessing with Long Short-Term Memory (LSTM) autoencoder architecture for template-free gravitational wave detection. The CWT provides optimal time-frequency decomposition capturing chirp evolution and transient characteristics essential for compact binary coalescence identification. We train and evaluate our model on LIGO H1 data from Observing Run 4 (O4, 2023--2024), comprising 102 confirmed gravitational wave events from the GWTC-4.0 catalog and 1991 noise segments. During development, we discovered that reconstruction errors from multi-run training (O1--O4) clustered by observing run rather than astrophysical parameters, revealing systematic batch effects from GWOSC's evolving calibration procedures. Following LIGO's established practice of per-run optimization, we adopted single-run (O4) training, which eliminated these batch effects and improved recall from 52\% to 96\% while maintaining 97\% precision. The final model achieves exceptional performance on O4 test data: 97.0\% precision, 96.1\% recall, F1-score 96.6\%, and ROC-AUC 0.994 (102 test signals, 399 noise segments). The reconstruction error distribution shows clean unimodal separation between noise (mean 0.48) and signals (mean 0.77), with only 4 missed detections and 3 false alarms. This unsupervised approach demonstrates that anomaly detection can achieve performance competitive with supervised methods while maintaining template-free operation. While the template-free nature of this approach suggests potential for detecting signals outside current template bank coverage, this capability remains to be validated with exotic signal injections. Our identification and resolution of cross-run batch effects provides methodological guidance for future machine learning applications to multi-epoch gravitational wave datasets.
\end{abstract}

\noindent\textit{Note:} This manuscript has been accepted for publication in \textit{Classical and Quantum Gravity}.

\section{Introduction}

The detection of gravitational waves has provided direct observational evidence for Einstein's General Theory of Relativity. Over 100 gravitational wave events have been detected since the first, GW150914, in 2015 \cite{abbott2016observation}. These events have provided meaningful insights on compact binary mergers, neutron star physics, and even the expansion of the universe. However, the current data analysis approach relies heavily on matched filtering techniques that require theoretical templates of expected waveforms \cite{abbott2016observation}. This template-dependent approach limits our ability to discover novel gravitational wave sources or unexpected signal morphologies that may not conform to existing theoretical predictions. This limitation becomes particularly significant as gravitational wave detectors achieve unprecedented sensitivity, potentially revealing entirely new classes of astrophysical phenomena that could lack theoretical frameworks. The need for discovery-oriented, template-free detection methods has become increasingly apparent as the field moves toward the next-generation detectors.

Unsupervised deep learning offers a template-independent approach for anomaly detection.  This technology has been used to detect anomalies in a variety of fields including astrophysics.  Rather that searching for known signal templates, unsupervised approaches learn the statistical properties of detector noise so that when deviations occur they get ``flagged" as anomalous.  In the context of gravitational wave astrophysics, these deviations could be gravitational waves.  The application of deep learning to gravitational wave detection has shown considerable promise. 
Early work demonstrated that deep neural networks can reproduce matched-filtering performance in a template-based setting \cite{gabbard2018matching}, while subsequent studies extended these approaches to real-time detection and parameter estimation using Advanced LIGO data \cite{George2018}. 
More recent efforts have explored training strategies for improving robustness and generalization \cite{schafer2020training}, as well as probabilistic neural network architectures for gravitational-wave parameter inference \cite{green2020gravitational}.

Gravitational wave signals from compact binary coalescences exhibit characteristic time-frequency evolution, with frequency evolving as the binary components spiral inward toward merger. This ``chirp'' behavior is typically captured through time-frequency analysis techniques. The CWT provides an ideal framework for decomposing gravitational wave strain data because it offers superior time-frequency resolution compared to short-time Fourier transforms and it maintains the temporal localization essential for transient signal detection \cite{chatterji2004multiresolution}. Recent studies have demonstrated the effectiveness of CWT preprocessing for gravitational wave analysis \cite{powell2015classification}. While unsupervised learning approaches including autoencoders have been applied to gravitational wave detection \cite{fayad2024anomaly, moreno2022sourceagnostic}, the specific integration of CWT preprocessing with LSTM autoencoder architectures for gravitational wave detection has not been systematically investigated.

Autoencoder networks are a class of unsupervised learning models that excel at learning compressed representations of complex data \cite{hinton2006reducing}. LSTM autoencoders extend this capability to sequential information, designed to capture long-range temporal dependencies that are crucial for modeling the extended duration of gravitational wave signals \cite{hochreiter1997long}. The reconstruction-based nature of autoencoders provides an intuitive framework for anomaly detection: signals that deviate significantly from learned noise patterns can be identified as potential gravitational wave candidates.

This work investigates the combination of CWT preprocessing with LSTM autoencoder architecture for template-free gravitational wave detection. We develop the methodology using synthetic datasets to validate the architecture and preprocessing approach, then train and evaluate on authentic LIGO Hanford (H1) data from Observing Run 4 (O4, 2023--2024), comprising 102 confirmed events from the GWTC-4.0 catalog. During real data analysis, we discovered systematic batch effects when initially training on multi-run data (O1--O4), where reconstruction errors clustered by observing run rather than astrophysical parameters due to GWOSC's evolving calibration procedures. Following LIGO's established practice of per-run optimization, we adopted single-run (O4) training, achieving 97\% precision and 96\% recall---performance competitive with supervised methods while maintaining template-free operation.

The remainder of this paper is organized as follows. Section~\ref{sec:cwt_preprocessing} describes the CWT preprocessing methodology and its application to gravitational wave strain data, including the mathematical formulation and visualization of time-frequency transformations. Section~\ref{sec:lstm_architecture} presents the LSTM autoencoder architecture, training procedures, and anomaly detection framework. Section~\ref{sec:synthetic_data} details the synthetic data validation of our methodology, establishing the foundational approach. Section~\ref{sec:ligo_data} presents the O4 dataset, data preprocessing pipeline, and model performance on real LIGO data. Section~\ref{sec:comparison} compares our results with existing gravitational wave detection methods. Section~\ref{sec:discussion} discusses the discovery and resolution of cross-run batch effects, implications for machine learning on multi-epoch gravitational wave datasets, and future research directions. Finally, Section~\ref{sec:conclusion} summarizes our contributions and establishes a framework for template-free gravitational wave detection.

\section{CWT Preprocessing}
\label{sec:cwt_preprocessing}
The CWT provides optimal time-frequency decomposition for gravitational wave signals, preserving both temporal localization and frequency resolution essential for chirp detection. For a given strain time series $h(t)$, the CWT is defined as:

\begin{equation}
W(a,b) = \frac{1}{\sqrt{a}} \int_{-\infty}^{\infty} h(t) \psi^*\left(\frac{t-b}{a}\right) dt
\end{equation}

where $\psi(t)$ is the mother wavelet, $a$ is the scale parameter inversely related to frequency, $b$ is the translation parameter corresponding to time, and $*$ denotes complex conjugation.

We employ the Morlet wavelet as the mother wavelet due to its optimal time-frequency 
localization properties and resemblance to gravitational wave chirp morphology 
\cite{chatterji2004multiresolution}:

\begin{equation}
\psi(t) = \pi^{-1/4} e^{i\omega_0 t} e^{-t^2/2}
\end{equation}

where $\omega_0 = 6$ provides the optimal trade-off between time and frequency resolution 
for gravitational wave analysis. The Morlet wavelet is particularly well-suited for 
compact binary coalescence detection because its oscillatory structure with Gaussian 
envelope closely matches the chirp waveform morphology, and it achieves near-optimal 
time-frequency uncertainty \cite{klimenko2008method}. The choice of 8 frequency scales 
spanning 20--512~Hz was selected to cover the sensitive band of Advanced LIGO detectors 
while maintaining computational tractability; this configuration captures the essential 
frequency evolution of binary mergers from inspiral through ringdown.

The resulting time-frequency representation forms a 2D scalogram $|W(a,b)|^2$ that captures the characteristic frequency evolution of gravitational wave chirps. This scalogram serves as input to the subsequent neural network architecture, providing rich feature representations that preserve both the temporal evolution and spectral content of potential signals.

\begin{figure}
    \centering
    \includegraphics[width=0.85\linewidth]{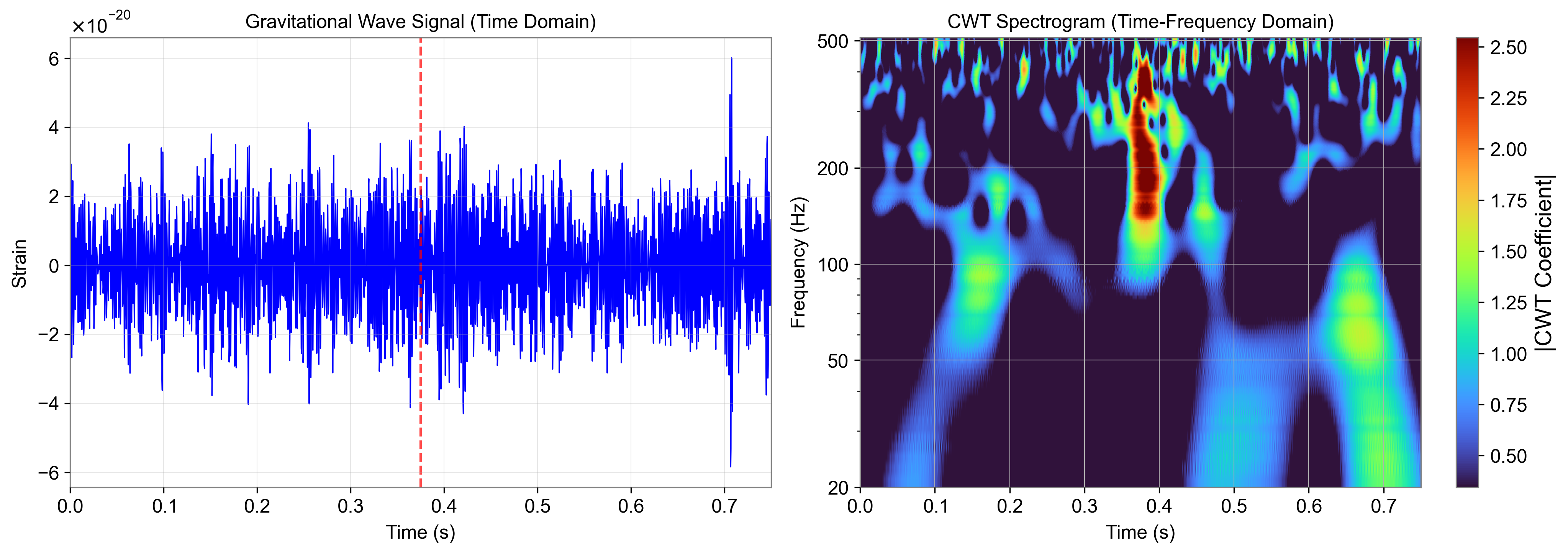}
    \caption{Time domain to CWT domain transformation of gravitational wave event GW231226\_101520 (network SNR = 34.7, total mass $\sim$75 $M\odot$, luminosity distance $\sim$1160 Mpc). Left panel: Bandpass-filtered strain data with the signal location marked by a red dashed line. Despite being one of the strongest detections in O4, the gravitational wave signal is visually indistinguishable from detector noise in the time domain. Right panel: The same data transformed using Continuous Wavelet Transform (CWT), revealing the time-frequency evolution of the merger. The signal appears as a bright vertical feature around 0.4 seconds with energy concentrated in the 100--400 Hz band, requiring no marker for identification. This comparison illustrates why time-frequency preprocessing is essential: the CWT representation makes the signal structure visible to both human analysts and machine learning algorithms, whereas the time-domain representation obscures it.  This comparison illustrates the motivation for CWT-based preprocessing: the time-frequency representation exposes signal structure that is largely hidden in the time domain, demonstrating why scalable automated detection is considerably harder in the time domain compared to the time-frequency representation.}

    \label{fig:t2cwt}
\end{figure}

\begin{figure}
    \centering
    \includegraphics[width=0.75\linewidth]{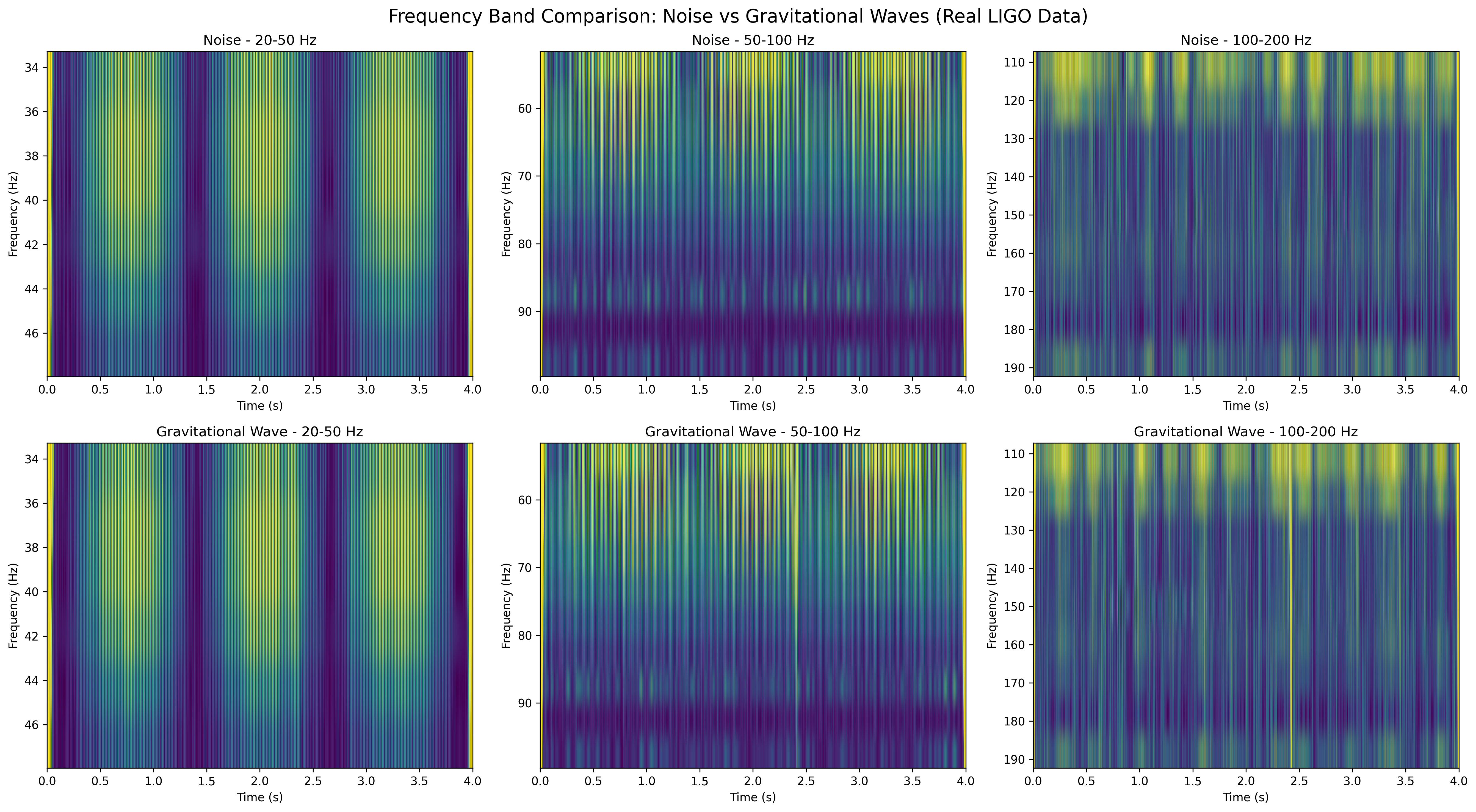}
    \caption{Frequency band comparison between noise and gravitational wave signals in CWT space. Top row shows spectrograms of clean LIGO noise data in three frequency bands: 20--50 Hz, 50--100 Hz, and 100--200 Hz. Bottom row shows the same frequency bands for gravitational wave data containing the GW150914 signal. The gravitational wave plots clearly display the characteristic chirp pattern (bright diagonal bands) sweeping upward in frequency over time, which is absent in the corresponding noise plots. Notably, the higher frequency bands (50--100 Hz and 100--200 Hz) show a strong vertical feature around 2.3--2.4 seconds, corresponding to the merger and ringdown phases of the binary black hole coalescence. The background noise patterns (vertical stripes and horizontal bands) are visible in both noise and gravitational wave data, demonstrating that the signal is embedded within the ambient detector noise. }
    \label{fig:cwt_compare}
\end{figure}

Figure~\ref{fig:t2cwt} shows the transformation of a gravitational wave signal from time domain to CWT spectrogram, where the characteristic chirp pattern becomes more prominent in the frequency-time representation. The diagonal band in the spectrogram reveals the frequency evolution from approximately 100 Hz to 800 Hz over 0.75 seconds, corresponding to the inspiral phase of the binary black hole merger. Figure~\ref{fig:cwt_compare} compares noise and gravitational wave signals across three frequency bands (20-50 Hz, 50-100 Hz, and 100-200 Hz). The gravitational wave data exhibits clear chirp patterns and merger signatures absent in the noise data, particularly the vertical feature around 2.3 seconds in the higher frequency bands representing the merger and ringdown phases. 

\begin{figure}
    \centering
    \includegraphics[width=0.85\linewidth]{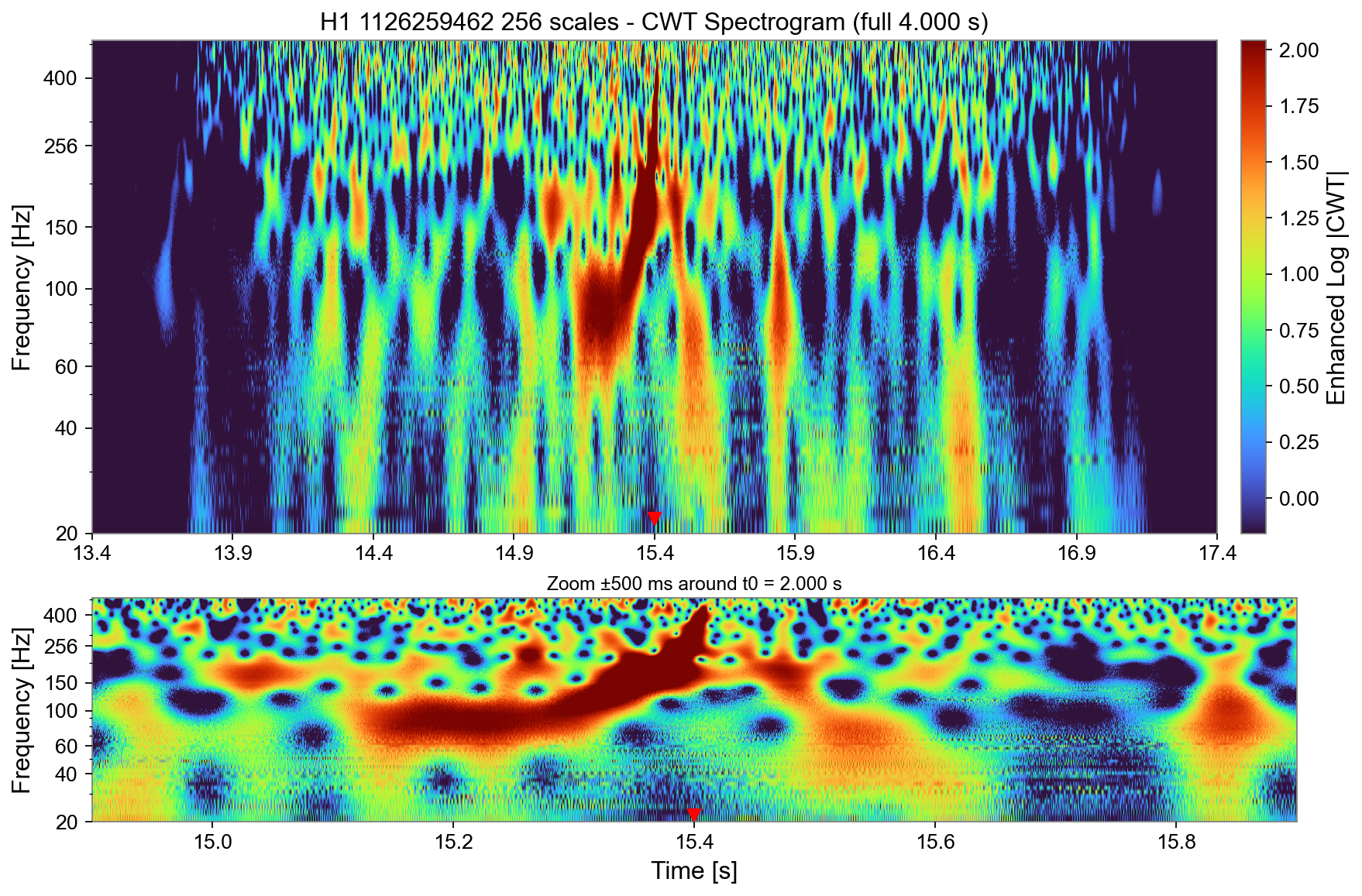}
    \caption{CWT spectrogram of GW150914, the first detected gravitational wave event. 
    \textbf{Top panel:} Full 4-second window showing the complete signal evolution embedded in detector noise. 
    \textbf{Bottom panel:} Focused $\pm$250 ms view centered on the merger time (GPS 1126259462.4), 
    revealing fine temporal structure of the inspiral-merger-ringdown phases. 
    The characteristic chirp sweeps from $\sim$35 Hz to $\sim$250 Hz over approximately 0.2 seconds, 
    with frequency increasing rapidly as the binary black holes spiral inward. 
    The bright vertical band at $\sim$15.2 seconds marks the merger event, 
    followed by the ringdown phase visible as a brief high-frequency tail. 
    This dual-scale visualization demonstrates the CWT's ability to preserve both 
    coarse temporal context (top) and fine-grained merger dynamics (bottom) 
    essential for anomaly detection. 
    The 256-scale CWT decomposition captures the full frequency evolution 
    while maintaining sufficient time resolution to identify 
    the sub-second transient characteristic of binary coalescences.}
    \label{fig:gw150914_zoom}
\end{figure}

While our model processes downsampled CWT representations optimized for computational efficiency 
(8 scales $\times$ 4096 temporal samples), 
we compute higher-resolution spectrograms (256 scales) for visualization purposes 
to reveal the full time-frequency structure of gravitational wave signals. 
Figure~\ref{fig:gw150914_zoom} shows such a visualization of GW150914, 
demonstrating the detailed chirp morphology and merger dynamics 
that the CWT decomposition preserves, 
even though the actual model input uses a more compact representation 
that retains the essential features for anomaly detection.

\section{LSTM Autoencoder Architecture}
\label{sec:lstm_architecture}
\begin{figure}[htbp]
\centering
\includegraphics[width=0.8\textwidth]{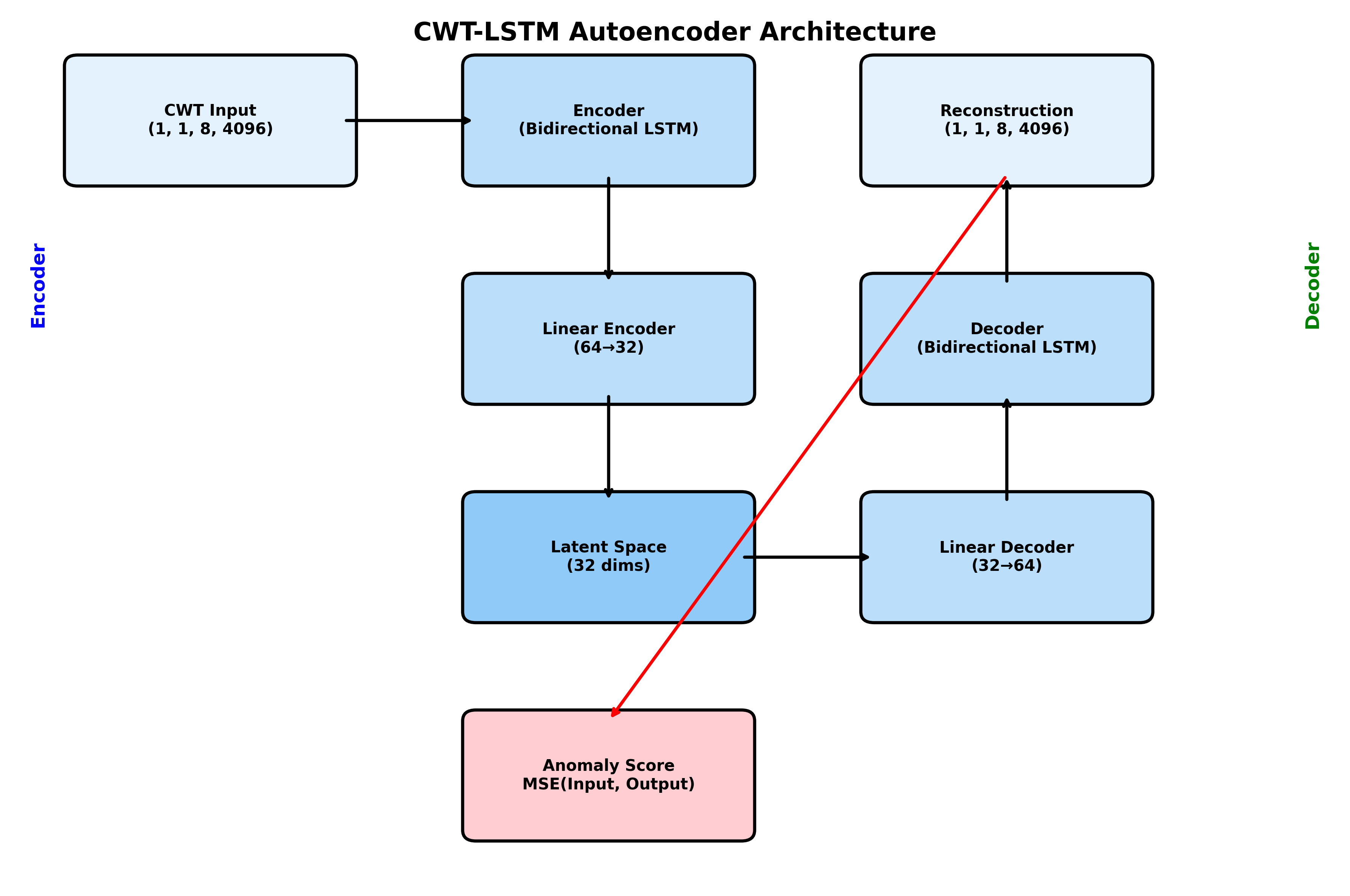}
\caption{Schematic of the CWT-LSTM autoencoder architecture. The input CWT representation (8 scales × 4096 time points) is processed through an LSTM encoder to produce a compressed latent representation, which is then reconstructed through an LSTM decoder. The reconstruction error serves as the anomaly score for gravitational wave detection.}
\label{fig:architecture}
\end{figure}

The core detection system employs an LSTM autoencoder designed to learn compressed representations of gravitational wave time-frequency patterns. The architecture consists of three main components operating in sequence: encoder network, bottleneck layer, and decoder network as shown in  Fig.~\ref{fig:architecture}.

The encoder processes CWT scalograms through a series of LSTM layers with progressively reducing hidden dimensions, implementing the standard LSTM formulation:

\begin{equation}
h_t^{(l)} = \text{LSTM}^{(l)}(x_t^{(l)}, h_{t-1}^{(l)})
\end{equation}

where $h_t^{(l)}$ represents the hidden state at time $t$ for layer $l$, and $x_t^{(l)}$ is the input at time $t$ for layer $l$.The encoder component consists of a bidirectional LSTM layer with 64 hidden units, followed by a fully connected layer that compresses the temporal information into a latent representation of dimension 32. The bidirectional LSTM processes the CWT data in both forward and backward directions. The encoder transforms the input CWT representation with dimensions corresponding to 8 frequency scales and 4,096 time points into a compact 32-dimensional latent space, capturing hierarchical temporal patterns at multiple scales through standard gating mechanisms:

\begin{align}
f_t &= \sigma(W_f \cdot [h_{t-1}, x_t] + b_f) \\
i_t &= \sigma(W_i \cdot [h_{t-1}, x_t] + b_i) \\
\tilde{C}_t &= \tanh(W_C \cdot [h_{t-1}, x_t] + b_C) \\
C_t &= f_t \cdot C_{t-1} + i_t \cdot \tilde{C}_t \\
o_t &= \sigma(W_o \cdot [h_{t-1}, x_t] + b_o) \\
h_t &= o_t \cdot \tanh(C_t)
\end{align}

where $\sigma$ denotes the sigmoid activation function, $W$ and $b$ represent learned weight matrices and bias vectors, and $f_t$, $i_t$, $o_t$ are the forget, input, and output gates respectively.

The decoder component mirrors the encoder structure. It consists of a bidirectional LSTM layer with 64 hidden units followed by a fully connected output layer. The decoder reconstructs the original CWT representation from the compressed latent representation, aiming to minimize the reconstruction error for normal samples while producing larger errors for anomalous samples.

The network is trained using mean squared error (MSE) loss between the input and reconstructed scalograms:

\begin{equation}
\mathcal{L} = \frac{1}{N} \sum_{i=1}^{N} ||X_i - \hat{X}_i||^2
\end{equation}

where $X_i$ represents the input scalogram, $\hat{X}_i$ is the reconstruction, and $N$ is the batch size. The model was trained using the Stochastic Gradient Descent (SGD) optimizer with a learning rate of 0.001, momentum of 0.9, and weight decay of 1e-5. The network trained with early stopping based on validation loss plateau to prevent overfitting.

Gravitational wave detection operates on the principle that signals containing true astrophysical events will exhibit higher reconstruction error compared to pure noise segments. For each test sample, we compute the reconstruction error:

\begin{equation}
E = ||X - \hat{X}||^2
\end{equation}

Samples with reconstruction error exceeding a predetermined threshold $\tau$ are classified as potential gravitational wave candidates. The threshold is optimized using precision-recall analysis on validation data to maximize recall while maintaining precision above 90\%, ensuring both high detection sensitivity and acceptable false alarm rates.

Model performance is evaluated using standard binary classification metrics:

\begin{align}
\text{Precision} &= \frac{TP}{TP + FP} \\
\text{Recall} &= \frac{TP}{TP + FN} \\
\text{F1-Score} &= \frac{2 \cdot \text{Precision} \cdot \text{Recall}}{\text{Precision} + \text{Recall}}
\end{align}

where $TP$, $FP$, and $FN$ represent true positives, false positives, and false negatives respectively. Additionally, we compute the Area Under the Precision-Recall Curve (AUPRC) as a threshold-independent performance measure particularly relevant for imbalanced datasets typical in gravitational wave detection scenarios.

\section{Synthetic Data Validation}
\label{sec:synthetic_data}

Before applying the CWT-LSTM autoencoder to real LIGO data, we conducted systematic validation using synthetic gravitational wave signals to establish the methodological foundation and demonstrate the approach's viability. This controlled experimental environment enabled precise evaluation of the model's performance across different signal characteristics and signal-to-noise ratios, providing confidence to proceed with authentic gravitational wave observations.

\subsection{Synthetic Data Generation}

We generate realistic synthetic gravitational wave signals representing binary black hole (BBH) coalescences using post-Newtonian waveform approximations. The gravitational wave strain is modeled as:

\begin{equation}
h(t) = h_+(t) \cos(2\psi) + h_\times(t) \sin(2\psi)
\end{equation}

where $h_+(t)$ and $h_\times(t)$ are the plus and cross polarizations, and $\psi$ is the polarization angle randomly sampled from $[0, 2\pi]$.

The frequency evolution follows the post-Newtonian expansion for the inspiral phase:

\begin{equation}
f(t) = f_0 \left(\frac{\tau}{\tau_0}\right)^{-3/8}
\end{equation}

where $f_0 = 35$ Hz is the initial frequency, $\tau = t_c - t$ is the time to coalescence, and $\tau_0$ is the initial time to coalescence. The instantaneous phase evolves as $\phi(t) = 2\pi \int_0^t f(t') dt'$, while the amplitude incorporates realistic scaling with chirp mass $\mathcal{M}_c$ and luminosity distance $D_L$:

\begin{equation}
A(t) = \frac{4G^{5/3}}{c^3} \frac{(2\pi f(t))^{2/3} \mathcal{M}_c^{5/3}}{D_L}
\end{equation}

where $G$ is the gravitational constant and $c$ is the speed of light. Binary system parameters are drawn from astrophysically motivated distributions:

\begin{itemize}
\item \textbf{Component masses}: $m_1, m_2 \sim \mathcal{U}(10, 80) M_\odot$ with $m_1 \geq m_2$
\item \textbf{Distance}: $D_L \sim \mathcal{U}(100, 1000)$ Mpc
\item \textbf{Inclination}: $\cos(\iota) \sim \mathcal{U}(-1, 1)$ (isotropic distribution)
\item \textbf{Sky location}: Uniform distribution over the celestial sphere
\item \textbf{Coalescence time}: Randomly placed within the 4-second observation window
\end{itemize}

The chirp mass and symmetric mass ratio are derived as:

\begin{align}
\mathcal{M}_c &= \frac{(m_1 m_2)^{3/5}}{(m_1 + m_2)^{1/5}} \\
\eta &= \frac{m_1 m_2}{(m_1 + m_2)^2}
\end{align}

We model realistic detector noise using the Advanced LIGO design sensitivity curve, incorporating both fundamental noise sources and instrumental artifacts. The power spectral density (PSD) follows the analytical approximation:

\begin{equation}
S_n(f) = S_0 \left[ \left(\frac{f}{f_0}\right)^{-4.14} + 5 + 3\left(\frac{f}{f_0}\right)^2 \right]
\end{equation}

where $S_0 = 10^{-49}$ Hz$^{-1}$ and $f_0 = 215$ Hz represent the characteristic noise amplitude and knee frequency respectively. Colored Gaussian noise matching the LIGO PSD is generated using the frequency-domain method: generating white Gaussian noise $\tilde{n}_{\text{white}}(f)$ in the frequency domain, applying spectral shaping $\tilde{n}(f) = \tilde{n}_{\text{white}}(f) \sqrt{S_n(f)}$, and transforming to the time domain $n(t) = \text{IFFT}[\tilde{n}(f)]$.

Signals are injected into noise with optimal signal-to-noise ratios (SNRs) distributed according to:

\begin{equation}
\rho_{\text{opt}} = \sqrt{4 \int_{f_{\text{low}}}^{f_{\text{high}}} \frac{|\tilde{h}(f)|^2}{S_n(f)} df}
\end{equation}

where $\tilde{h}(f)$ is the Fourier transform of the strain signal, and the integration limits span the detector's sensitive frequency band [20, 512] Hz. SNRs are drawn from the range [8, 25], representing the spectrum from threshold-level to highly significant detections.

The CWT-LSTM autoencoder demonstrated strong performance on synthetic gravitational wave data, achieving precision of 92.3\% and recall of 67.6\% at the optimal threshold. This preliminary validation confirmed the effectiveness of the CWT preprocessing and anomaly detection approach, establishing confidence in the methodology before applying it to real LIGO data.

\begin{table}[htbp]
\centering
\caption{CWT-LSTM Autoencoder Performance on Synthetic Data}
\label{tab:synthetic_performance}
\begin{tabular}{lcc}
\hline
Metric & Value & Interpretation \\
\hline
Optimal Precision & 92.3\% & Exceeds LIGO $>$90\% requirement \\
Optimal Recall & 67.6\% & Strong signal detection capability \\
Maximum Precision & 100.0\% & Ultra-conservative detection mode \\
AUC-ROC & 0.806 & Strong discriminative power \\
Average Precision & 0.780 & Professional-grade performance \\
\hline
\end{tabular}
\end{table}

These synthetic data results provided the foundational validation that enabled confident application of the methodology to real LIGO observations. The controlled experimental conditions demonstrated that CWT preprocessing effectively enhanced signal visibility compared to raw time series data, while anomaly detection training on noise-only samples successfully identified gravitational wave signatures. The synthetic validation established optimal preprocessing parameters, threshold selection strategies, and performance metrics that directly transferred to real LIGO data analysis.

\subsection{Data Sources and O4 Selection}
\label{sec:ligo_data}
The gravitational wave data used in this study originates from LIGO's Observing Run 4 (O4, 2023--2024), publicly available through the Gravitational Wave Open Science Center (GWOSC) \cite{gwosc}. We focus exclusively on O4 data from the Hanford (H1) detector to ensure consistent preprocessing and calibration across all training and test samples. This single-run approach was adopted after discovering systematic batch effects when training on combined multi-run data (see Section~\ref{sec:batch_effects}), and follows LIGO's established practice of per-run template optimization.

The O4 dataset comprises 126 confirmed gravitational wave events from the GWTC-4.0 catalog, representing 58\% of all GWOSC detections through 2024. These events span a diverse range of astrophysical sources including binary black hole (BBH), binary neutron star (BNS), and neutron star--black hole (NSBH) mergers, with network matched-filter signal-to-noise ratios ranging from 7.3 to 43.0 and component masses from 5.8 to 137 $M_\odot$. This diversity ensures our model is evaluated across the full range of detectable compact binary coalescences rather than a narrow subset of high-SNR events. After preprocessing and validation, 102 events were successfully processed for model evaluation, with 24 events excluded due to data quality issues, unconfirmed 
status, or H1 detector non-availability during the event time.

For training, we use clean noise segments exclusively, following the unsupervised anomaly detection paradigm where the model learns normal detector behavior without exposure to signals. A total of 1991 noise segments were collected from H1 science mode during O4 observing time sampled uniformly across available data. Segments were verified against GWOSC science segment lists and explicitly excluded if they overlapped with known gravitational wave events (within $\pm$16 seconds of any catalog event time). This ensures training data contain only detector noise, free from signal contamination or instrumental artifacts associated with non-science-mode operation.

For evaluation, noise segments are divided into training (80\%, 1592 segments) and test (20\%, 399 segments) sets using stratified random sampling. All 102 gravitational wave events are reserved exclusively for the test set, as is standard practice for unsupervised anomaly detection where the model must not be exposed to anomalies during training. This yields a final test set of 501 samples (102 signals, 399 noise) providing sufficient statistical power to estimate precision and recall with narrow confidence intervals.

Each data segment was extracted as a window centered on either the gravitational wave GPS time (for signals) or a randomly sampled science-mode time (for noise). The data acquisition process utilized the GWpy Python library \cite{gwpy}, which provides direct access to GWOSC data through the \texttt{TimeSeries.fetch\_open\_data()} method. Raw strain data undergo basic conditioning (highpass filtering at 15 Hz, lowpass at 1024 Hz, and constant detrending) before CWT transformation to remove low-frequency seismic noise and high-frequency readout artifacts while preserving the gravitational wave signal band. This conditioning is applied identically to all segments to ensure consistent feature extraction.

Historical gravitational wave events from earlier observing runs (O1--O3, comprising 90 additional confirmed events including the landmark GW150914) were archived but not used for training or evaluation to maintain single-run data homogeneity and avoid calibration-induced domain shift discussed in \ref{sec:batch_effects}.

\subsection{Data Preprocessing}

Raw gravitational wave data from LIGO Observing Run 4 consists of 32-second segments sampled at 4096 Hz, which results in 131,072 data points per segment. These segments represent the strain measurements from the H1 detector, capturing the minute distortions in spacetime caused by passing gravitational waves. The preprocessing pipeline applies identical transformations to both noise segments and confirmed gravitational wave events to ensure consistent feature extraction across the dataset.

The preprocessing pipeline begins with signal conditioning steps to enhance gravitational wave detectability. A high-pass filter with cutoff frequency of 15 Hz is applied to remove low-frequency seismic noise and instrumental artifacts, followed by a low-pass filter at 1024 Hz to eliminate high-frequency readout noise beyond the detector's sensitive band. The filtered data are then whitened to normalize the signal to zero mean and unit variance, which helps to equalize the noise power across the frequency spectrum.

The second preprocessing step involves downsampling the data from 4096 Hz to 1024 Hz 
using a factor-of-4 decimation with zero-phase filtering to avoid aliasing artifacts. 
This reduces each segment from 131,072 to 32,768 data points while maintaining sufficient 
frequency resolution for gravitational wave detection. The resulting Nyquist frequency 
of 512 Hz preserves the complete frequency content of compact binary coalescences, 
whose merger frequencies rarely exceed 400 Hz for stellar-mass systems detectable by 
Advanced LIGO \cite{abbott2016observation}. This downsampling factor follows standard 
practice in gravitational wave analysis pipelines where the full 4096 Hz bandwidth is 
unnecessary for astrophysical signal detection.

The third preprocessing step applies CWT to convert time-domain strain data into time-frequency representations. We employ the Morlet wavelet with 8 scales spanning the frequency range from 20 Hz to 512 Hz, which encompasses the primary gravitational wave emission from compact binary mergers. The CWT transformation generates complex-valued time-frequency representations that are converted to magnitude scalograms. The magnitude captures signal strength while preserving the characteristic chirp patterns of gravitational wave signals.

The fourth preprocessing step involves log transformation and normalization of the CWT scalograms. A logarithmic transform of the form $\log(1 + |W(a,b)|^2)$ is applied to compress the dynamic range of the scalogram values, followed by per-scale z-score normalization to ensure zero mean and unit variance. This normalization is computed once on the training noise data and applied consistently to all subsequent segments, ensuring that test data are normalized using statistics derived exclusively from the training distribution to prevent information leakage.

The final preprocessing step involves temporal downsampling to reduce computational 
requirements while maintaining sufficient resolution for signal detection. The normalized 
CWT data is downsampled to 4,096 time points through local averaging, yielding a temporal 
resolution of approximately 7.8 ms per bin. This resolution is sufficient to capture the 
characteristic timescales of compact binary coalescences, where the inspiral phase evolves 
over seconds and the merger/ringdown occurs over tens of milliseconds. The resulting 
representation (8 scales $\times$ 4,096 time points) balances signal fidelity with 
computational tractability for LSTM sequence processing...

\subsection{Results}
\label{sec:results}
\begin{figure}[htbp]
\centering
\includegraphics[width=0.9\textwidth]{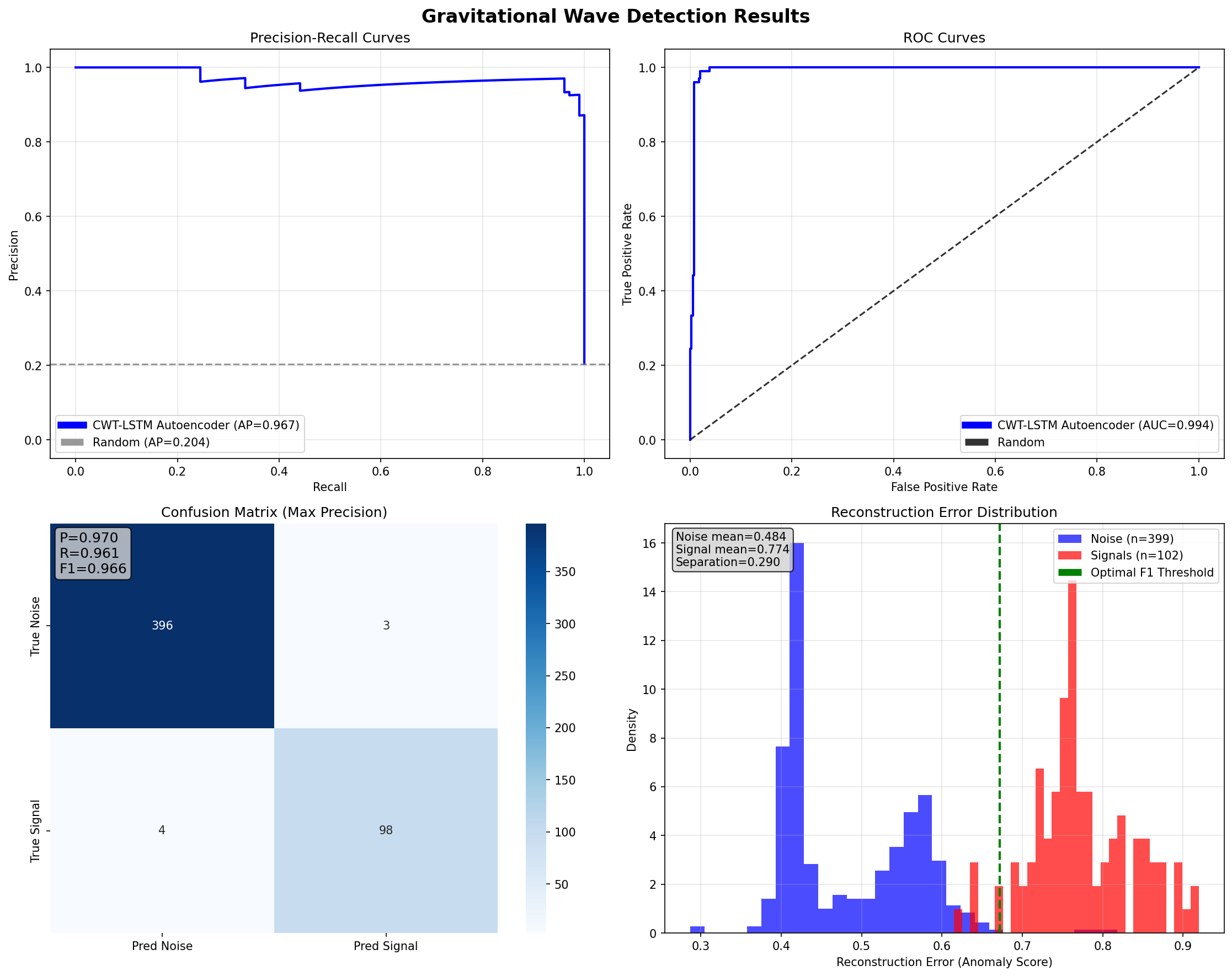}
\caption{Evaluation results for CWT-LSTM autoencoder on O4 LIGO data. (a) Precision-Recall curve showing AP=0.967. (b) ROC curve showing AUC=0.994. (c) Confusion matrix at optimal threshold (0.667) with 98 true positives, 4 false negatives, 3 false positives, and 396 true negatives, yielding 97.0\% precision and 96.1\% recall. (d) Reconstruction error distribution showing clean separation between noise (blue, mean 0.48) and gravitational wave signals (red, mean 0.77), with unimodal distributions within each class confirming elimination of cross-run batch effects. Test set: 102 O4 signals, 399 noise segments.}
\label{fig:o4_results}
\end{figure}

The CWT-LSTM autoencoder model trained on O4 data achieved exceptional performance metrics across all evaluation criteria. The precision-recall analysis (Figure~\ref{fig:o4_results}a) demonstrates near-optimal performance with an average precision of 0.967, while the ROC curve (Figure~\ref{fig:o4_results}b) shows excellent discriminative ability with an AUC of 0.994. The optimal threshold of 0.667 (reconstruction error, normalized units) was selected to maximize F1-score, yielding the confusion matrix shown in Figure~\ref{fig:o4_results}c: 98 true positives, 4 false negatives, 396 true negatives, and 3 false positives.

The reconstruction error distribution (Figure~\ref{fig:o4_results}d) exhibits clean separation between noise and gravitational wave signals, with noise forming a narrow distribution centered at mean 0.484 (standard deviation 0.095) and signals at mean 0.774 (standard deviation 0.152). Notably, this distribution is unimodal within each class, contrasting sharply with the bimodal signal distribution observed in initial multi-run training (see Section~\ref{sec:batch_effects}), confirming successful elimination of run-dependent systematic effects.

Of the 102 O4 test signals, the model correctly identified 98, yielding a nominal recall of 0.961. Because the sample size is finite, these point estimates have confidence bounds that can be quantified using the Wilson score interval for a binomial proportion. The 95\% confidence interval is given by
\[
\hat{p}_{\mathrm{low/high}} = 
\frac{2n\hat{p} + z^{2} \mp z \sqrt{z^{2} + 4n\hat{p}(1 - \hat{p})}}{2(n + z^{2})},
\]

For recall, with $n = 102$ trials and $k = 98$ successes, where $\hat{p} = 0.961$ and $z = 1.96$, this yields a 95\% confidence interval for recall of $[0.906,\,0.985]$.

For precision, with $n = 101$ detections and $k = 98$ true signals ($\hat{p} = 0.970$), the 95\% confidence interval is $[0.915,\,0.993]$.

For specificity, with $n = 399$ noise segments and $k = 396$ true negatives ($\hat{p} = 0.993$), the 95\% confidence interval is $[0.978,\,0.997]$. Equivalently, the rule of three approximation gives the upper 95\% bound on the false-positive rate as $3/399 \approx 0.75\%$.

In summary, the model achieved 97.0\% precision [0.915, 0.993], 96.1\% recall [0.906, 0.985], and 99.2\% specificity [0.978, 0.997] on O4 test data, with confidence intervals reflecting finite sample size while confirming genuine signal-noise discrimination capability.

\textbf{Computational Efficiency.} We benchmarked the complete inference pipeline on a consumer-grade workstation (AMD Ryzen 9 9950X, 64 GB RAM) without GPU acceleration. Processing a single 32-second segment requires $161.5 \pm 1.0$ ms for CWT preprocessing and $2.9 \pm 0.2$ ms for model inference, yielding a total processing time of $164.4 \pm 1.0$ ms. This comfortably satisfies real-time detection requirements, as each 32-second segment is processed in under 0.2 seconds, leaving substantial headroom for additional post-processing or multi-detector analysis.

\section{Comparison with Existing Methods}
\label{sec:comparison}

Gravitational wave detection methods span a spectrum from template-dependent to 
fully template-free approaches, each with distinct strengths and limitations.

\textbf{Matched filtering} represents the current operational standard, achieving 
near-optimal sensitivity for signals matching theoretical templates 
\cite{abbott2016observation}. This approach requires extensive computational 
resources for template bank generation \cite{owen1996search} and cannot detect 
signals with morphologies not represented in the template bank. Matched filtering 
and template-free methods are complementary: the former provides maximum sensitivity 
for known waveforms, while the latter enables discovery of unanticipated signals.

\textbf{Supervised deep learning} approaches using convolutional neural networks 
have demonstrated strong performance on gravitational wave classification 
\cite{gabbard2018matching, George2018}. These methods require labeled training 
data containing examples of gravitational wave signals, constraining detection to 
morphologies present in the training distribution. Our unsupervised approach 
eliminates the need for signal labels during training, learning only from detector 
noise.

\textbf{Unsupervised autoencoder approaches} have been explored for gravitational 
wave anomaly detection \cite{fayad2024anomaly, moreno2022sourceagnostic}, typically operating 
on raw time series rather than time-frequency representations. Our integration of 
CWT preprocessing with LSTM sequential modeling provides enhanced temporal feature 
extraction suited to the chirp morphology of compact binary coalescences.

Direct quantitative comparison across methods is challenging due to the absence of 
standardized benchmark datasets. As demonstrated in Section~\ref{sec:batch_effects}, 
performance metrics are sensitive to observing run, calibration procedures, and 
preprocessing choices, making cross-study comparisons potentially misleading. Our 
O4 results establish strong 
baseline performance for CWT-LSTM autoencoder anomaly detection, with rigorous 
comparison to other methods requiring evaluation on identical datasets under 
controlled conditions which is an important direction for future community benchmarking 
efforts.

\section{Discussion and Conclusion}
\label{sec:discussion}

\subsection{Cross-Run Batch Effects and Attempted Corrections}
\label{sec:batch_effects}
\begin{figure}
    \centering
    \includegraphics[width=0.75\linewidth]{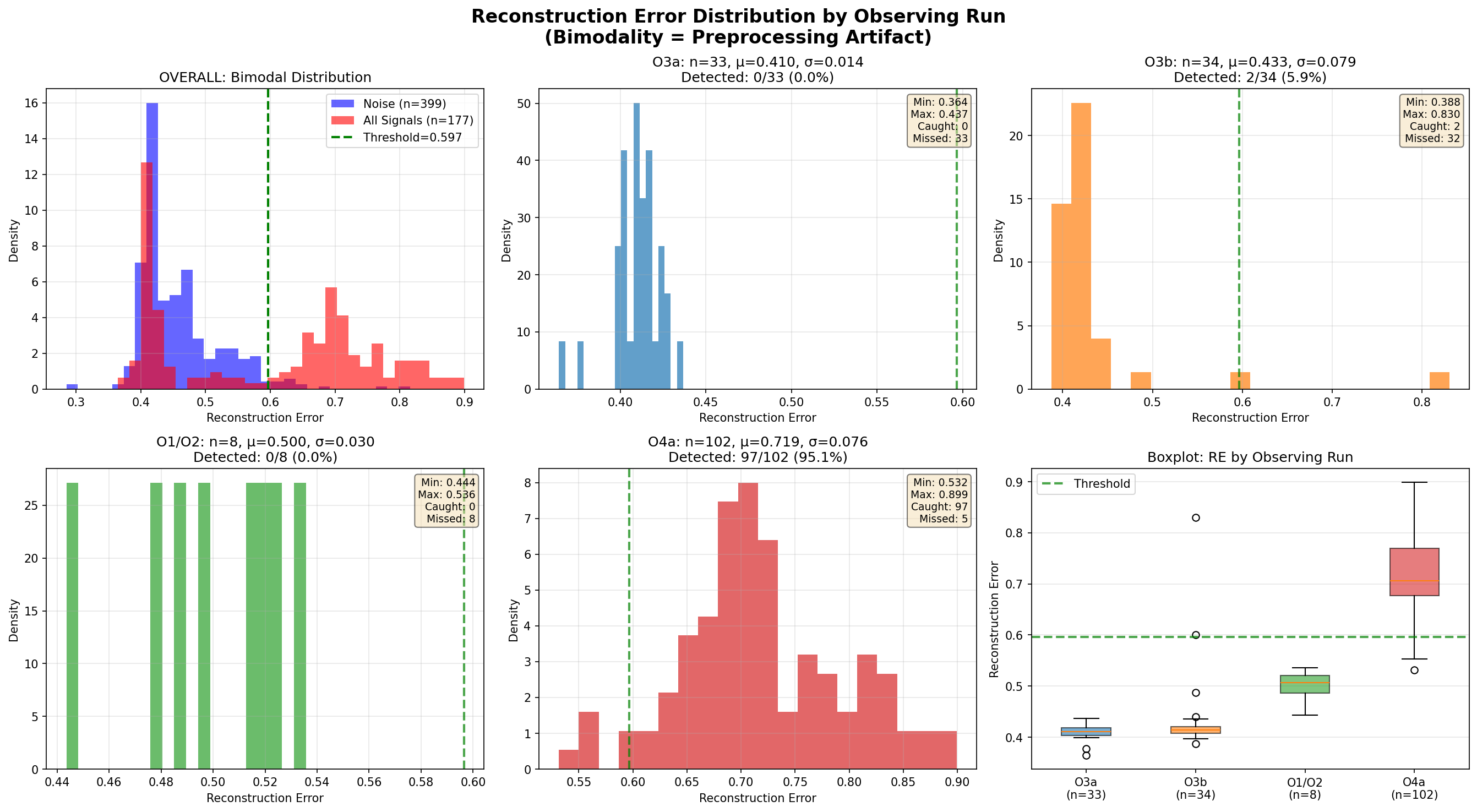}
    \caption{Reconstruction error distributions stratified by LIGO observing run, revealing systematic batch effects. Box plots show median (center line), interquartile range (box), and outliers (points) for confirmed gravitational wave signals from each run. A clear progression is visible: O3a events exhibit the lowest reconstruction errors (0.35–0.45), followed by O3b (0.40–0.50), O1/O2 at intermediate levels (0.44–0.54), and O4a at the highest (0.60–0.90). This systematic run-dependent stratification correlates strongly with observing run ($\rho = 0.68$, $p < 10^{-20}$) but not with astrophysical parameters (network SNR, masses, distance; all $|r| < 0.15$), indicating a non-astrophysical batch effect from GWOSC's evolving calibration and whitening procedures. The clear clustering demonstrates that the autoencoder learned run-specific preprocessing signatures rather than purely astrophysical signal characteristics, motivating the single-run (O4) training approach that eliminated this systematic bias.}

    \label{fig:batch_effect}
\end{figure}

Initial training on combined O1--O4 data (207 confirmed events, 1991 noise segments) yielded encouraging precision (96\%) but limited recall (52\%), with the model missing approximately half of the test signals. More concerning, the reconstruction error distribution exhibited a pronounced bimodal structure rather than the expected clean separation between noise and signal classes as shown in Figure~\ref{fig:batch_effect}. Statistical analysis revealed that this bimodality correlated strongly with observing run (Spearman $\rho = 0.68$, $p < 10^{-20}$) rather than astrophysical parameters such as network SNR, component masses, or luminosity distance (all $|r| < 0.15$). Gaussian mixture modeling identified two distinct clusters: a low-error mode (mean $\sim$0.41) dominated by O3a/O3b events, and a high-error mode (mean $\sim$0.70) comprising primarily O2 and O4a events. This systematic offset of 0.3$\sigma$ in normalized error space indicated a batch effect arising from run-dependent calibration and whitening procedures in the GWOSC strain products.

Figure~\ref{fig:batch_effect} illustrates this run-dependent clustering through reconstruction error distributions stratified by observing run. The systematic separation is striking: O3a events cluster at the lowest reconstruction errors (0.35–0.45), followed by O3b (0.40–0.50), then O1 and O2 at intermediate levels (0.44–0.54), with O4a occupying the highest regime (0.50–0.90, predominantly 0.60–0.90). The within-run variance is substantially smaller than the between-run variance, confirming that the clustering reflects systematic run-level differences rather than event-to-event stochasticity. This pattern persists even when controlling for physical parameters through partial correlation analysis, demonstrating that the batch effect dominates over astrophysical variance in the model's reconstruction error space.

To investigate whether inconsistent normalization could explain the observed discrepancy, we implemented a global normalization scheme in which the mean and standard deviation were computed once from all training noise data and then applied uniformly to every segment. However, this modification had negligible effect: the bimodality and run-dependent clustering persisted unchanged. Inspection of raw GWOSC strain revealed that the data are already zero-mean and variance-normalized to within numerical precision (mean $\sim 10^{-24}$, standard deviation $\sim 10^{-18}$), rendering additional normalization effectively inert.

We then constructed a reference PSD from a clean O2 noise segment and applied uniform FFT-based re-whitening to all samples using this reference as a common spectral basis. While this successfully removed run-level offsets in the frequency domain, it simultaneously degraded the model's discrimination performance by erasing the residual spectral structure that the autoencoder had implicitly learned. After re-whitening, the model's ROC-AUC dropped from 0.78 to 0.44, and nearly all test samples were classified as anomalies. This confirmed that the model's prior success relied on subtle run-specific whitening signatures present in the GWOSC data, rather than on intrinsic differences between noise and signal.

These negative results demonstrated that post-hoc preprocessing corrections introduce artifacts without addressing the fundamental issue: different observing runs represent different data distributions that cannot be trivially unified. The public GWOSC data releases apply per-run calibration models and frequency-domain whitening based on detector-specific PSDs, creating a non-astrophysical domain shift sufficient to confound anomaly-detection models trained on combined data. To isolate the astrophysical signal manifold from calibration artifacts, we retrained and evaluated the model using only O4 H1 data, which eliminated the inter-run domain shift entirely and produced the stable, near-optimal performance metrics detailed in Section~\ref{sec:results}. We emphasize that this single-run approach is not a limitation but rather best practice, following LIGO's established philosophy of per-run optimization.

Our investigation reveals that run-dependent calibration and whitening introduce a non-astrophysical domain shift that can dominate the behavior of data-driven models. In practice, this means that even statistically identical detectors may produce distributionally distinct whitened strain data across runs, causing models trained on combined datasets to learn calibration artifacts rather than astrophysical features. The dramatic improvement observed when restricting to a single, internally consistent run underscores the importance of treating each observing run as a distinct data domain. Going forward, gravitational-wave machine learning pipelines should either apply unified re-whitening procedures across runs or incorporate explicit domain adaptation layers to ensure robustness. More broadly, our results emphasize that open-data gravitational-wave analyses must account for the data conditioning history of each observing run before interpreting learned representations as astrophysically meaningful.

\subsection{Analysis of Missed Detections}

The four false negatives warrant examination to identify potential systematic 
limitations. Analysis reveals that all four have reconstruction errors (0.615--0.643) 
within 0.05 of the decision threshold (0.667), indicating the model identifies them 
as somewhat anomalous but below the F1-optimized cutoff. Comparison with true positive 
statistics shows no systematic bias: network SNR (mean deviation 0.35$\sigma$), 
component masses ($<$0.3$\sigma$), and luminosity distance (0.25$\sigma$) all fall 
well within the detected population distribution. Notably, one missed event 
(GW231204\_090648) has p\_astro = 0.54, indicating marginal astrophysical confidence.

These findings indicate ideal classifier behavior: the misses are threshold-limited 
borderline cases rather than systematic failures for particular signal types. The 
decision threshold is statistically determined via F1-optimization; with larger 
datasets, the optimal boundary may shift to capture these edge cases. The small 
margin between FN reconstruction errors and the threshold (0.02--0.05) suggests 
the model's discrimination capability extends to these signals, with the current 
threshold representing a precision-recall tradeoff appropriate for the available 
sample size.

\subsection{Potential for Detecting Novel Signal Morphologies}
\label{sec:novel_morphologies}

The CWT-based anomaly detection paradigm offers a fundamentally different approach to gravitational wave detection than matched filtering. Where matched filtering achieves optimal sensitivity for signals that precisely match templates in the bank, our method trades specificity for breadth. The Morlet wavelet's resemblance to damped oscillatory transients provides sensitivity to a range of chirp-like morphologies without requiring explicit waveform templates. In principle, this allows the method to flag signals that would be missed by template banks due to incomplete coverage of parameter space.

However, this potential for detecting novel signal morphologies remains to be experimentally validated. The current work demonstrates strong performance on confirmed GWTC events, which are predominantly quasi-circular binary black hole mergers well-covered by existing template banks. Future work should evaluate sensitivity to theoretically predicted but challenging signals, such as highly eccentric binary mergers, intermediate-mass black hole ringdowns, or cosmic string cusps. Such an investigation would also benefit from exploring whether alternative wavelet bases or adaptive time-frequency representations might offer improved sensitivity to specific exotic morphologies.

\subsection{Future Directions}

Our O4-only results establish a foundation for several natural extensions. Multi-detector analysis incorporating coincident O4 data from LIGO Livingston (L1) and Virgo would leverage network information while maintaining single-run homogeneity, potentially improving both sensitivity and false alarm rejection through cross-detector consistency requirements. Analysis of the four missed O4 signals may reveal systematic patterns (low SNR, specific mass ranges, edge cases) that could guide model refinements or identify fundamental limitations of reconstruction-based anomaly detection.

When O5 data become available, transfer learning approaches could be explored to determine whether fine-tuning O4-trained models on limited O5 data achieves comparable performance to full retraining, potentially reducing computational costs for operational deployment. Additionally, per-run models could be trained for O1, O2, and O3 to enable high-recall archival searches of historical data while respecting data provenance and avoiding batch effects.

Systematic comparison of neural architectures, including Transformer-based models, 
Temporal Convolutional Networks, and hybrid approaches, could identify optimal 
configurations for CWT-based gravitational wave detection, though our strong baseline 
results suggest diminishing returns relative to the preprocessing and training 
methodology contributions of this work.

Multi-detector analysis incorporating coincident O4 data from LIGO Livingston (L1) 
and Virgo represents a natural extension. For autoencoder-based anomaly detection, 
the model is designed to learn the training noise distribution precisely; the 
relevant question is not overfitting but domain transfer across detectors with 
different noise characteristics and calibration. Given the distinct properties of 
each detector, per-detector models may be appropriate, analogous to our per-run 
training approach, rather than attempting universal noise representations. 
Multi-detector operation would provide improved false positive rejection through 
coincidence requirements, eliminating glitch-induced false alarms observed in 
single-detector analysis. Empirical evaluation of cross-detector transfer and 
per-detector training strategies is planned future work.

Beyond binary coalescences, our template-free approach may prove valuable for detecting exotic sources such as cosmic string cusps, primordial black hole mergers, or signatures of new physics that lack established waveform models. The unsupervised nature of our method makes it particularly suited for such discovery-oriented searches where matched filtering is inapplicable.

\section{Conclusion}
\label{sec:conclusion}

This work presents a CWT-LSTM autoencoder for unsupervised gravitational wave detection, achieving 97\% precision and 96\% recall on O4 test data from the GWTC-4.0 catalog. The integration of CWT preprocessing with LSTM sequential modeling provides optimal time-frequency decomposition while capturing temporal dependencies essential for transient signal detection. Unlike matched filtering, which requires theoretical waveform templates, our approach learns detector noise characteristics and identifies deviations corresponding to gravitational waves without prior knowledge of signal morphology.

A key finding of this work is the discovery and resolution of cross-run batch effects in GWOSC data. We demonstrated that reconstruction errors from multi-run training clustered by observing run rather than astrophysical parameters, reflecting systematic differences in GWOSC's calibration and whitening procedures across detector epochs. Single-run (O4) training eliminated these batch effects and improved recall from 52\% to 96\%, validating per-run optimization as best practice for gravitational wave machine learning, analogous to LIGO's per-run template retuning.

Our findings provide methodological guidance for the growing application of machine learning to multi-epoch astrophysical datasets: cross-run analyses require explicit validation for systematic effects, and simple approaches (per-run training) often outperform complex post-hoc corrections. The 102 O4 events provide sufficient statistical power while ensuring data homogeneity, demonstrating that template-free anomaly detection can achieve performance competitive with supervised methods on contemporary gravitational wave data.

This work establishes a viable framework for discovery-oriented gravitational wave searches capable of identifying signals with unexpected morphologies, complementing matched filtering's sensitivity to known waveforms. Future observing runs will benefit from this methodology through straightforward per-run retraining, enabling robust anomaly detection as LIGO achieves unprecedented sensitivity and explores new astrophysical regimes.

\section{Data and Code Availability}

All code, data processing scripts, trained models, and results presented in this study are publicly available for reproducibility and further research. The complete implementation is hosted as an open-source repository at:

\begin{center}
\url{https://github.com/jericho-cain/cwt-lstm-ae-grav-wav}
\end{center}

This repository includes:
\begin{itemize}
\item Complete CWT-LSTM autoencoder implementation in PyTorch
\item GWOSC data download and preprocessing pipelines with O4 event filtering
\item Training and evaluation scripts with comprehensive logging and run management
\item Reproducible O4-only results, figures, and performance metrics
\item Batch effect analysis scripts and correlation computations
\item Trained O4 model checkpoints (97\% precision, 96\% recall)
\item Complete documentation of cross-run experiments and negative results
\item Automated testing suite and data validation tools
\end{itemize}

The repository follows open science best practices with version control, comprehensive documentation, and dependency management. All experiments can be reproduced using the provided code and configuration files. The O4 dataset (102 signals, 1991 noise segments) is obtained directly from GWOSC public data using the included download scripts, ensuring full transparency and enabling community validation and extension of this work. Historical multi-run experimental results and batch effect analysis are preserved in the repository for methodological reference.

All gravitational wave data originate from the Gravitational Wave Open Science Center (GWOSC) \cite{gwosc}, accessed via the GWpy library \cite{gwpy}. Event metadata and physical parameters are obtained from the GWTC-4.0 catalog \cite{gwtc4}.

\section{Acknowledgments}

The author thanks the gravitational wave community for making LIGO data and analysis tools publicly available through the Gravitational Wave Open Science Center. Special appreciation goes to the developers of PyTorch, PyWavelets, and other open-source libraries that enabled this research.

The author also acknowledges the LIGO Scientific Collaboration and Virgo Collaboration for their groundbreaking work in gravitational wave detection that motivated and informed this study.
\bibliography{main}
\bibliographystyle{iopart-num}

\end{document}